\documentclass[12pt,onecolumn]{emulateapj}
\input epsf

\begin{document}
\title{Measuring the redshift dependence of the CMB monopole temperature with PLANCK data.} 
\author{I. de Martino\altaffilmark{1}, F. Atrio-Barandela\altaffilmark{1}, A. da Silva\altaffilmark{2}, H. Ebeling\altaffilmark{3}, A. Kashlinsky\altaffilmark{4}, D. Kocevski\altaffilmark{5}, C.J.A.P. Martins\altaffilmark{2}}
\altaffiltext{1}{F\'{\i}sica Te\'orica, Universidad de Salamanca, 37008 Salamanca, Spain;
email: ivan.demartino@usal.es; atrio@usal.es}
\altaffiltext{2}{Centro de Astrofisica da Universidade do Porto, Rua das Estrelas s/n,
4150-762 Porto, Portugal; email: asilva@astro.up.pt; Carlos.Martins@astro.up.pt}
\altaffiltext{3}{Institute for Astronomy, University of Hawaii, 2680 Woodlawn Drive, Honolulu, HI 96822, USA; email: ebeling@ifa.hawaii.edu}
\altaffiltext{4}{SSAI and Observational Cosmology Laboratory, Code 665, Goddard Space Flight Center, Greenbelt, MD 20771, USA alexander.kashlinsky@nasa.gov}
\altaffiltext{5}{Department of Physics, University of California at Davis, 1 Shields Avenue, Davis, CA 95616, USA; email: kocevski@physics.ucdavis.edu}

\begin{abstract}
We study the power of PLANCK data to constrain deviations of the
Cosmic Microwave Background black body temperature from adiabatic evolution
using the thermal Sunyaev-Zeldovich anisotropy induced by clusters of galaxies.
We consider two types of data sets: the cosmological signal is removed
in the Time Ordered Information or is removed from the final maps; and
two different statistical estimators, based on the ratio of temperature anisotropies
at two different frequencies and on a fit to the spectral variation of
the cluster signal with frequency. To test for systematics,
we construct a template from clusters drawn from a hydro-simulation
included in the pre-launch Planck Sky Model. We demonstrate that, using a 
proprietary catalog of X-ray selected clusters with measured redshifts,
electron densities and X-ray temperatures, we can constrain  deviations
of adiabatic evolution, measured by the parameter
$\alpha$ in the redshift scaling $T(z)=T_0(1+z)^{1-\alpha}$, with
an accuracy of $\sigma_\alpha=0.011$ in the most optimal case 
and with $\sigma_\alpha=0.016$ for a less optimal case.
These results represent a factor 2-3 improvement 
over similar measurements carried out using  quasar spectral lines
and a factor 6-20 with respect to earlier results using smaller cluster samples. 
\end{abstract}

\keywords{Cosmic Microwave Background. Cosmology: theory. Cosmology: observations}

\maketitle

\bigskip
\section{Introduction}

One of the fundamental tenets of the Big-Bang paradigm
is the adiabatic evolution of the Universe. Early thermal equilibrium
among the different particle species, entropy and photon number conservation
produce a Cosmic Microwave Background (CMB) with a blackbody spectrum.
The CMB temperature was measured to be $T_0=2.725\pm 0.002$K
by the Far Infrared Absolute Spectrometer (FIRAS) of the Cosmic Background
Explorer (COBE) satellite (Mather et al 1999). The adiabatic expansion
of the Universe and photon number conservation imply that the CMB temperature 
evolves with redshift as $T(z)=T_0(1+z)$. Establishing observationally this 
relation would test our current understanding of the Universe since
models like decaying vacuum energy density and gravitational `adiabatic' photon 
creation predict different scaling relations (Overduin \& 
Cooperstock, 1998; Matyjasek 1995; Lima et al 2000; Puy 2004;
Jetzer et al 2011, Jetzer \& Tortora 2012). In these models, energy is slowly 
injected without producing distortions on the blackbody spectrum, 
evading the tight FIRAS constraints. Nevertheless, in these models
the blackbody temperature scales nonlinearly as $T(z)=T_0(1+z)^{1-\alpha}$.
Therefore, measuring the redshift dependence of the CMB black body temperature at
various cosmological epochs can provide strong constraints on physical
theories at the fundamental level. 

There are currently two methods to determine $T(z)$ at redshifts $z>0$:
(a) using fine structure lines from interstellar atoms or molecules,
present in quasar spectra, whose transition energies are 
excited by the CMB photon bath (Bahcall \& Wolfe, 1968) and (b) 
from the thermal Sunyaev-Zeldovich anisotropies (hereafter
TSZ, Sunyaev \& Zel'dovich, 1972, 1980) due to the inverse Compton scattering 
of photons by the free electrons within the potential wells of 
clusters of galaxies.
Early observations of fine-structure levels of atomic
species like carbon only led to upper limits on $T(z)$ because
the CMB is not the only radiation field populating the energy
levels and collisional excitation is an important contribution.
Assuming the CMB is the only source of excitation, 
Songaila et al. (1994) measured $T(z=1.776)=7.4\pm 0.8$K; but
collisional excitation was not negligible and it had to be corrected.
The first unambiguous measurement was only achieved six years later, with
a considerably larger error bar (Srianand et al. 2000).
Lately, Noterdaeme et al (2010) 
succeeded in obtaining a direct and precise measurements from the rotational 
excitation of CO molecules. They constrained the deviation
from linear scaling to be $\alpha=-0.007\pm 0.027$ at $z\sim 3$.
Battistelli et al. (2002) reported the first observations of $T(z)$
using the TSZ effect of the COMA and A2163 clusters of galaxies
with $\alpha=-0.16^{+0.34}_{-0.32}$.
Luzzi et al. (2009) determined the CMB temperature in the redshift
range $z=0.023-0.546$, from the measurements of 13 clusters.
They restricted their analysis to $\alpha\in[0,1]$ and set up an
upper limit of $\alpha\le 0.079$ at the 68\% confidence level.
No significant deviations from the redshift dependence of the CMB temperature
predicted in the standard model has been found.

While there is interest in doing such observations as 
far back as possible (which one can do with spectroscopic methods), 
low-redshift measurements play an important role too.  First, 
the two techniques are complementary with each other since they have different
systematics and probe the adiabatic evolution of the Universe at different redshifts.
Spectroscopic observations probe the matter era, roughly between redshifts
$z=2-4$, while TSZ probes the epoch of dark energy domination,
$z\le 1$. In particular, these measurements can shed light on the
onset of dark energy domination; in many models this is associated with 
as a phase transition (Mortonson et al. 2009, Nunes et al. 2009)
which could leave imprints in the $T(z)$ relation. Second,
in models where photon number is not conserved, not only does 
the temperature-redshift relation change,
but so does the distance duality relation (Etherington 1933), and these
departures from the standard behavior are not independent. This link
between the two relations requires information at all redshifts
and will, when better datasets become available,
be a powerful consistency test for the standard cosmological paradigm 
(Avgoustidis et al 2012).

The Planck mission has been designed to produce a full-sky survey
of the CMB with unprecedented accuracy in temperature and polarization
(Planck Collaboration 2011a). The instrument operates at nine frequencies
logarithmically spaced in the range 30-857GHz. The in-flight
performance of the High and Low Frequency Instruments have been 
described by the Planck HFI Core Team (2011a) and Mennella et al (2011). 
Due to its large frequency coverage, high resolution and low noise, 
is an optimal instrument for blind detection of clusters using the 
TSZ effect. The first clusters detected by PLANCK
include 189 cluster candidates with signal-to-noise larger than 6
(Planck Collaboration 2011b). These SZ clusters are mostly at moderate redshifts 
(86\% had $z<0.3$) and span over a decade in mass, up to the rarest and most 
massive clusters with masses above $10^{15}$M$_\odot$.
In this article we analyze how PLANCK data can be used to test the 
standard scaling relation of the CMB temperature with redshift.
We use a two-fold approach: first, our pipeline is tested on simulated clusters
drawn from a full hydrodynamical simulation; second, using a catalog of
$623$ clusters derived from ROSAT data and with well measured X-ray 
properties, we predict the accuracy that PLANCK measurements will
reach using those clusters. 
In comparison with earlier analysis of Horellou et al (2005), we use a catalog
of X-ray selected clusters and in our simulations, gas evolution is fully 
taken into account. Briefly, in Sec 2 we describe our methodology; in Sec 3
we discuss our data and simulations; in Sec 4 we explain our pipeline; 
in Sec 5 we present our results and in Sec 6 we summarize our main conclusions. 
The final goal of the paper is to forecast the accuracy with which PLANCK 
will constraint $\alpha$ for our cluster sample. 

\begin{center}
\begin{table}[!h]
\begin{tabular}{|l|rrrrrr|}
\hline
Planck Channel & 1 & 2 & 3 & 4 & 5 & 6 \\
\hline
Central frequency $\nu_0$/GHz & 44 &70 &100 &143 &217 &353\\
Frequency resolution $\Delta\nu$ (FWHM/GHz) & 8.8 & 14 &33 &47 &72 &116 \\
Angular resolution $\Delta\theta$ (FWHM/arcmin) & 26.8 & 13.1 & 9.8& 7.1 & 5.0& 5.0\\
Noise per pixel $\sigma_{noise}/\mu$K (Blue Book) & 51 &  52 & 15 & 12 & 19 &  58\\
Noise per pixel $\sigma_{noise}/\mu$K (in-flight performance)&
109 &  96 &  14 &  9 & 13 & 49 \\
\hline
\end{tabular}
\caption{Technical details of Planck channels used in this study.
The noise per pixel in-flight performance corresponds to 1 year of integration.}
\end{table}
\end{center}

\bigskip
\section{Methodology.}

Compton scattering of CMB photons by the hot Intra-Cluster (IC) gas induces 
secondary temperature anisotropies on the CMB radiation in the direction 
of clusters of galaxies. There are two components: the thermal (TSZ,
Sunyaev \& Zeldovich 1972) due to the thermal motion of the IC medium with 
temperature $T_e$ and the kinematic (KSZ, Sunyaev \& Zeldovich 1980)
due to the motion of the cluster with speed $\vec{v}_{cl}$
respect to the isotropic CMB frame. Neglecting relativistic
corrections, the TSZ and KSZ contributions to the temperature
anisotropy in the direction of a cluster $\hat{n}$ are given by
\begin{equation}
\frac{T(\hat{n})-T_0}{T_0}=\int \left[G(\nu)\frac{k_BT_e}{m_ec^2}+
\frac{\vec{v}_{cl}\hat{n}}{c}\right]d\tau=G(\nu)y_c+\tau\frac{\vec{v}_{cl}\hat{n}}{c}
\label{eq:sz}
\end{equation}
In this expression, $d\tau=\sigma_Tn_edl$ is the cluster optical depth and
$n_e(l)$ the electron density evaluated along the line of sight $l$, $\sigma_T$
is Thomson cross section,
$T_0$ the current CMB mean temperature, $k_B$ the Boltzmann constant, $m_ec^2$ the 
electron annihilation temperature, $c$ the speed of light and $\nu$ the frequency
of observation. The Comptonization parameter is defined 
as $y_c=(k_B\sigma_T/m_ec^2)\int n_eT_e dl$.
Due to its frequency dependence $G(\nu)$, the TSZ is a distortion of the CMB spectrum.
Its amplitude is independent on the cluster distance, making
it a useful tool to detect clusters at high redshifts. All known astrophysical 
foregrounds have a different dependence with frequency so clusters can be
clearly detected in CMB maps with enough frequency coverage. 
In the non-relativistic limit, $G(x)= x{\rm coth}(x/2)-4$.
The reduced frequency $x$ is given by $x=h\nu(z)/kT(z)$ with
$\nu(z)$ the frequency of a CMB photon scattered off by the IC gas and
$T(z)$ the black body temperature of the CMB at the cluster location.

If the Universe evolves adiabatically, $T(z)=T_0(1+z)$. Due to the
expansion, the frequency of a photon scattered by the
IC plasma at redshift $z$ is Doppler shifted as: $\nu(z)=\nu_0(1+z)$
and the ratio $x=h\nu(z)/kT(z)=h\nu_0/kT_0=x_0$ is
independent of redshift. If the evolution of the Universe is non-adiabatic, 
the temperature-redshift relation would not be constant.  
Two functional forms have been considered in the literature:
$T(z)=T_0(1+z)^{1-\alpha}$  (Lima et al 2000) and 
$T(z)=T_0(1+bz)$ (LoSecco et al 2001). In both cases, the 
photon frequency is assumed to be redshifted as in the standard model:
$\nu(z)=\nu_0(1+z)$. Since the largest fraction of known clusters
of galaxies are at redshifts below $z\le 0.7-1$, the differences
between both redshift dependences are small so we will 
only analyze the first model. The reduced frequency varies as 
$x=x_0(1+z)^\alpha$ and 
the spectral frequency dependence of the TSZ effect, $G(\nu)$ now
depends on $\alpha$: $G(x)=G(\nu,\alpha)$. 
Using the TSZ effect, two methods have been proposed to constraint $\alpha$. 
Fabbri et al. (1978) proposed to measure 
the zero cross frequency of clusters at different redshifts
that, for adiabatic evolution, occurs at $\nu\simeq 217GHz$.
Rephaeli (1980) suggested to use the ratio of the TSZ anisotropy at 
different scales, $R(\nu_1,\nu_2,\alpha)= G(\nu_1,\alpha)/G(\nu_2,\alpha)$. 
Both methods have different systematics.
By taking ratios, the dependence on the Comptonization parameter 
is removed and the need to account for model uncertainties on the gas density
and temperature profile is avoided. At the same time, the analysis is 
more complicated since the distribution of temperature anisotropy ratios is 
highly non-gaussian (Luzzi et al. 2009). The measurement of the cross over
frequency is also problematic since the TSZ is inherently weak and could
be dominated by uncertain systematics. For this reason, the measurements
carried out thus far (Battistelli et al 2002, Luzzi et al 2009),
based in a small number of clusters, have concentrated
in the ratio method. As an alternative to the zero frequency method,
we will fit the TSZ signal at different frequencies and we will measure
the function $G(\nu,\alpha)$. We shall denote this procedure the {\it Fit Method}.

\begin{figure}
\centering 
\epsfxsize=.495\textwidth \epsfbox{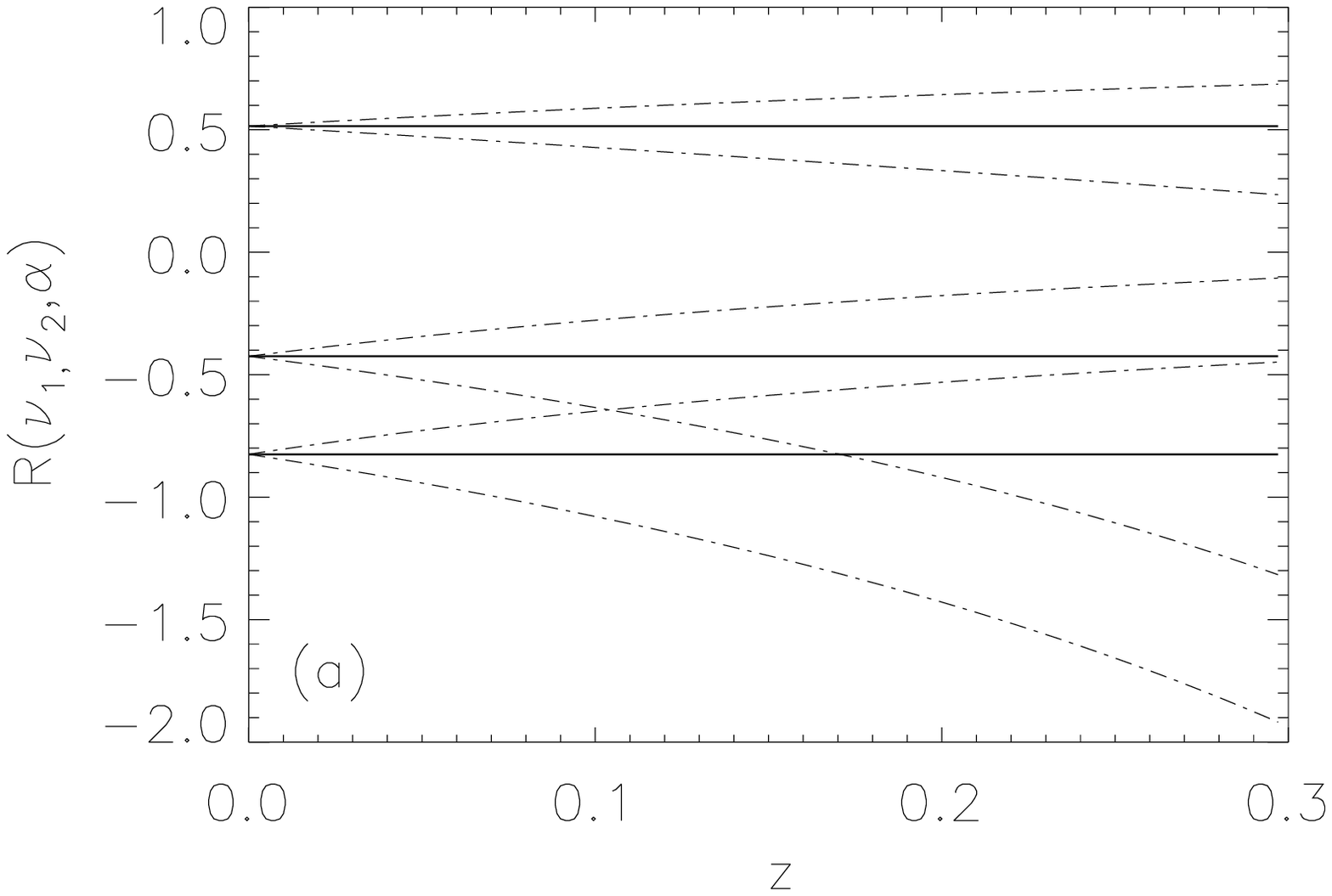}
\epsfxsize=.495\textwidth \epsfbox{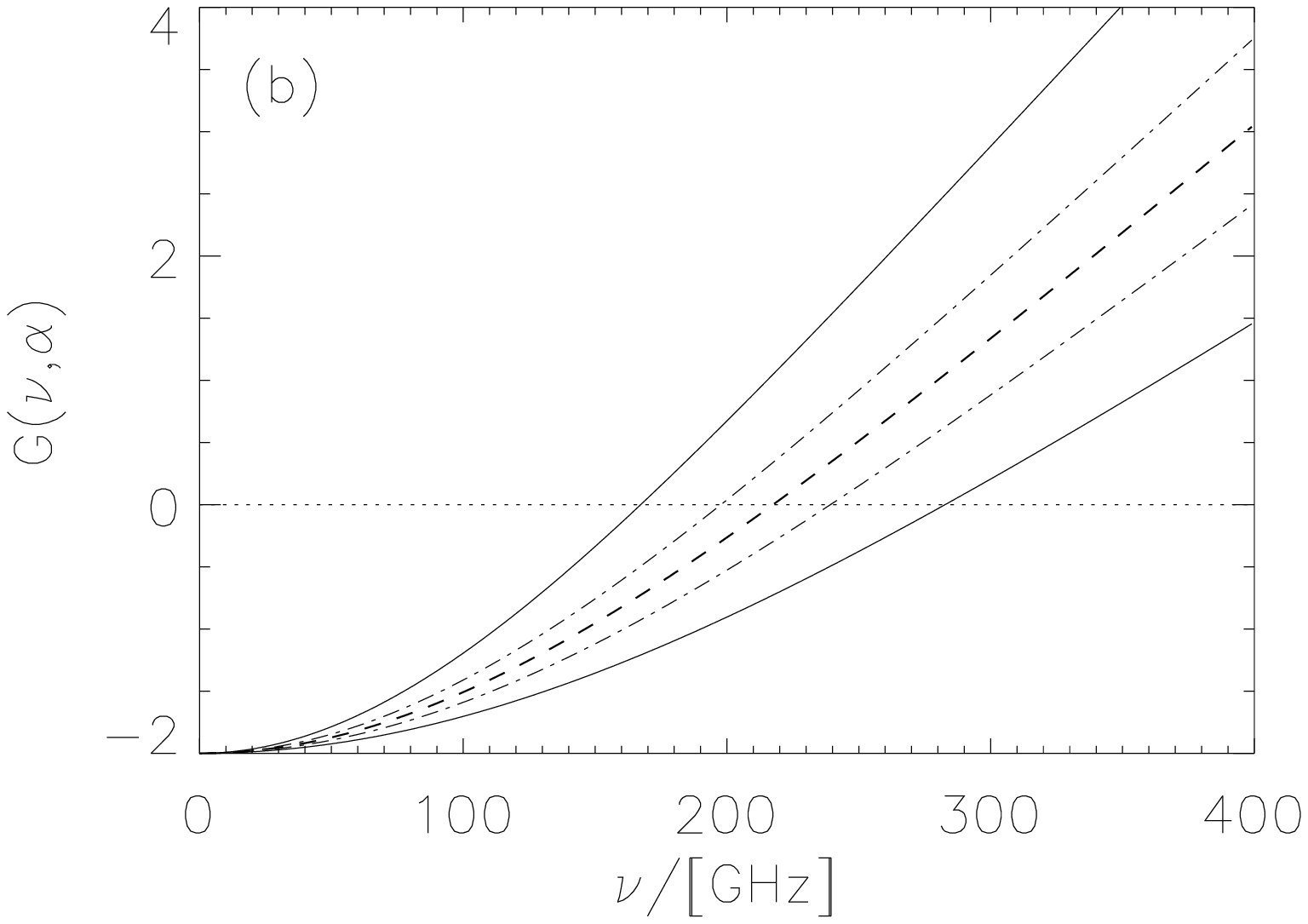}
\caption{(a) variation of the ratio $R(\nu,353GHz,\alpha)$ as a function of redshift
for $\nu=143$GHz (top set of curves), 100GHz (middle set) and 44GHz (lower set).
The solid straight line corresponds to adiabatic evolution $\alpha=0$ and
the dot-dashed lines represent $\alpha=-1,1$.
(b) Spectral dependence of the TSZ effect $G(\nu,\alpha)$ for 
two clusters located at two clusters located at z=0.3 (solid lines) and 
z=0.1 (dot-dashed lines) with $\alpha=-1,1$. The dashed line corresponds
to adiabatic evolution $\alpha=0$, that is identical for any redshift.
The zero cross frequency occurs when those lines cross the dotted line.}
\label{fig1}
\end{figure}

The function $G(\nu,\alpha)$ characterizes uniquely the TSZ contribution. 
At each frequency, PLANCK Low Frequency Instrument (LFI) receivers and 
High Frequency Instrument (HFI) bolometers are sensitive to a wide range of 
frequencies and the spectral dependence is not $G(\nu)$ but 
\begin{equation}
\bar{G}(\nu_0,\alpha)=\int_0^\infty G(\nu,\alpha)e^{(\nu-\nu_0)^2/2\sigma_\nu^2} d\nu
\label{eq:freq}
\end{equation}
Hereafter, we will remove the upper bar and $G(\nu,\alpha)$ will refer 
to the averaged frequency dependence of eq~(\ref{eq:freq}).
In Fig~\ref{fig1} we plot the temperature ratio (Fig~\ref{fig1}a)
and frequency dependence (Fig~\ref{fig1}b) for different values of $\alpha$. 
In Fig~\ref{fig1}a, the solid line represents the adiabatic evolution model
$\alpha=0$ that is independent of redshift; the dot-dashed lines bound
the region where $\alpha=-1,1$. From top to bottom, the
ratios are $R(\nu,353GHz,\alpha)$ with $\nu=143,100,44$GHz.
In Fig~\ref{fig1}b, we plot the spectral dependence $G(\nu,\alpha)$ for 
adiabatic evolution ($\alpha=0$, dashed line) and $\alpha=-1,1$ for a cluster at 
redshift $z=0.1$ (dot-dashed line) and $z=0.3$ (solid line).
The null TSZ signal, represented by the dotted line, shows that the
zero cross frequency varies in the range $\nu\sim 170-270$GHz.

To construct a pipeline that implements the ratio or zero cross frequency
tests we need to consider the specifics of the PLANCK data.
The cosmological CMB signal is the dominant 
contribution except at the most massive clusters.
Foreground residuals or astrophysical contaminants,
while smaller in amplitude, would induce systematic shifts in 
the Comptonization parameter (Aghanim, Hansen \& Lagache, 2005) varying
with frequency and biasing the redshift dependence of the TSZ effect.
To characterize the noise and foreground emission, the HFI and LFI core teams
have constructed maps with the CMB cosmological contribution subtracted off the Time 
Ordered Information (TOI). They have used six different component separation 
algorithms to remove the primordial CMB signal. The difference 
between the six methods provides an estimate the CMB residual 
(Planck HFI Core Team 2011b, Zacchei et al 2011). 
The resulting maps are dominated by noise and foreground residuals. Due to the
scanning strategy, the noise is rather inhomogeneous, largely dominated by a
white noise component plus a $1/f$ contribution. In the LFI instrument,
the $1/f$ noise is largest at 30GHz. For the HFI, the noise is largest at 545 
and 857GHz. Also, those channels have the smallest resolution and we will
not be considered in this work. Later we shall show that this is not
a limitation since the channels with the highest resolution are the ones
with the largest statistical power to determine $\alpha$.
The technical details of the maps considered in this study are listed in Table 1: 
central frequency and FWHM of the antenna spectral response function, approximated by
a gaussian with FWHM $\Delta\nu$, angular resolution and noise per pixel. 
We indicate the required Blue Book 
specifications\footnote{http://www.rssd.esa.int/SA/PLANCK/docs/Bluebook-ESA-SCI(2005)1\_V2.pdf}) 
and the in-flight measured noise per pixel after one year of integration
(Mennella et al 2011, Planck HFI Core Team 2011a). 
With respect to astrophysical contaminants,
WMAP used the K, Ka Differencing Assemblies 
and the extinction corrected H$\alpha$
maps (Finkbeiner 2003) to subtract the synchrotron and free-free
emissions and the Finkbeiner et al (1999) map to subtract the dust contribution
(see Gold et al 2009 for details). A similar analysis using 
the 30, 545 and 857 GHz channels, and foreground templates could 
be used to remove the foreground contribution. Ideally, one would
obtain maps without cosmological signal and with a low level of
foreground residuals.

\begin{figure}
\centering 
\epsfxsize=.495\textwidth \epsfbox{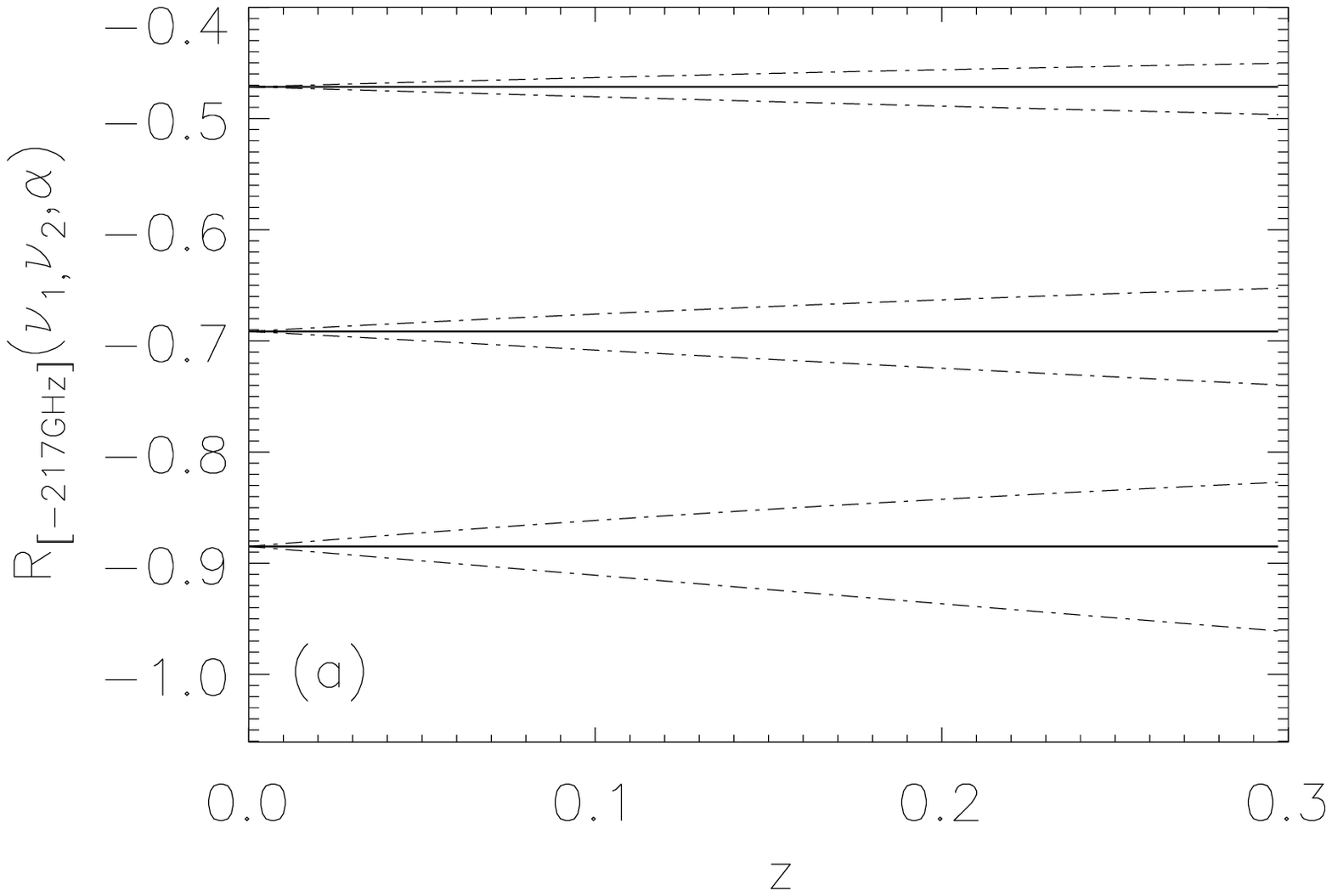}
\epsfxsize=.495\textwidth \epsfbox{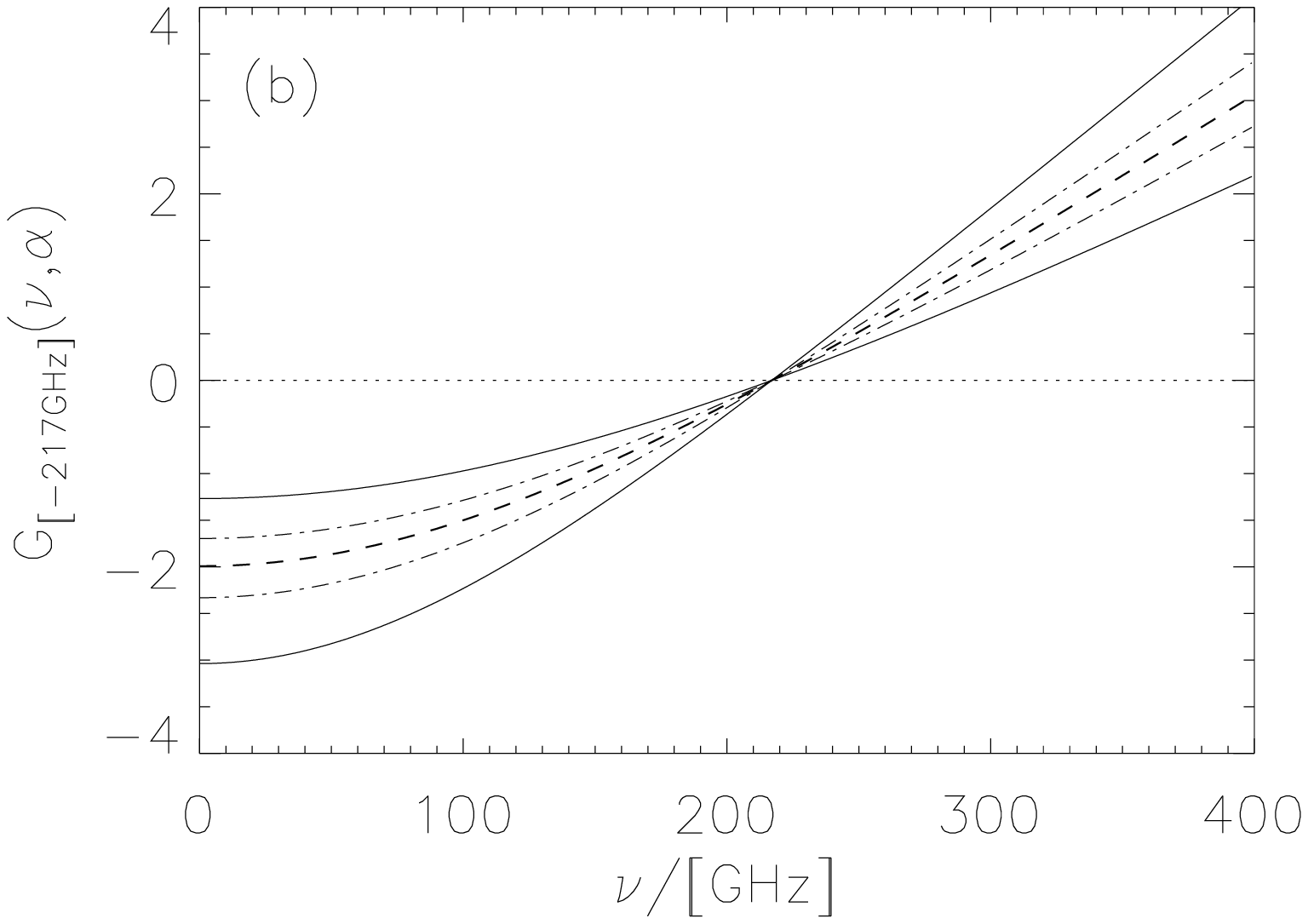}
\caption{(a) Ratio $R_{[-217GHz]}(\nu,353GHz,\alpha)$ for $\nu=143,100$ and $44$GHz, 
and (b) Spectral dependence of $G(\nu,\alpha)-G(217GHz,\alpha)$. 
Curves follow the same convention as in Fig.~\ref{fig1}.}
\label{fig2}
\end{figure}

Maps free of the intrinsic CMB signal and/or foreground residuals
are the most convenient to test the adiabatic evolution of the Universe. 
Removing the cosmological CMB signal on the TOI data leaves an unknown level
of CMB residuals, whose distribution and power spectrum are difficult to 
model and the final results could be biased in an unmeasurable way.
Alternatively, the intrinsic CMB can be removed by subtracting the highest resolution
map, conveniently degraded, from the other maps. Both techniques have
different systematics and the consistency of the results would be a test
of their validity. Therefore, we shall carry out two type of simulations,
depending on what data sets become available:
(A) CMB subtraction in the TOI plus foreground removal using templates
would produce maps with only instrumental noise, KSZ and TSZ
with some unknown levels of primordial CMB and foreground residuals. 
Deviations from adiabatic evolution can be measured by taking ratios
of temperature anisotropies at different frequencies (Fig~\ref{fig1}a)
or by fitting the spectral dependence $G(\alpha,\nu)$ (Fig~\ref{fig1}b) 
to the data. (B) The component separation takes place on the final maps. 
The primordial CMB is removed using the foreground clean 217GHz map.
We degrade the angular resolution of the 217GHz channel to that
of the other 5 channels before subtracting it from the corresponding map.
We checked that the intrinsic CMB and KSZ anisotropies are 
removed exactly but the frequency dependence of the TSZ effect is modified.
In Fig~\ref{fig2}a, we plot the ratio of the CMB-KSZ removed
maps at different frequencies: 
$R_{[-217GHz]}(\nu_1,\nu_2,\alpha)=[G(\nu_1,\alpha)-G(217GHz,\alpha)]/
[G(\nu_2,\alpha)-G(217GHz,\alpha)]$. In Fig~\ref{fig2}b we represent 
$G_{[-217GHz]}(\nu,\alpha)=G(\nu,\alpha)-G(217GHz,\alpha)$.
The lines follow the same conventions than in Fig.~\ref{fig1}. 

Our pipeline will analyze maps using CMB subtraction on the TOI
(method A) and on the maps themselves (method B). We will
compute $\alpha$ with both the ratio and the fit method. 
By subtracting the 217GHz, the estimator $R_{[-217GHz]}$
has a much weaker dependence on $\alpha$ than $R$.
With respect to the fit method, 
we do not measure $G(\nu,\alpha)$ directly but 
$\Delta  T(\hat{n})=T_0y_cG(\nu,\alpha)$. To measure 
the spectral shape from the data we need an independent 
determination of $y_c$ using X-ray data. We will 
use a proprietary catalog of X-ray selected clusters 
with all the required information (see Sec.~\ref{sec:xray})
and for this catalog we will forecast the constrain to be achieved
with PLANCK.

\bigskip
\bigskip
\section{Cluster templates and final maps.}

To test our systematics,
we will construct two TSZ templates: one based on our catalog
of X-ray selected clusters, the other based on an all-sky hydrodynamical simulation. 
The final maps were simulated using the HealPix package (Gorski et al 2005)
with resolution $N_{side}=1024$. 

\subsection{Y-map from X-ray selected clusters. \label{sec:xray}}

Our cluster sample contains 623 clusters outside WMAP Kp0 mask.
It was created combining the
ROSAT-ESO Flux Limited X-ray catalog (REFLEX, B\"ohringer et al 2004)
in the southern hemisphere, the extended Brightest Cluster Sample
(eBCS, Ebeling et al 1998, Ebeling et al 2000) in the north, and the
Clusters in the Zone of Avoidance (CIZA, Ebeling, Mullis \& Tully
2002, Kocevski et al 2007) sample along the Galactic plane.  All three
surveys are X-ray selected and X-ray flux limited. A detailed description 
of the creation of the merged catalog is given in Kocevski \& Ebeling (2006). 
The position, flux, X-ray luminosity and angular extent of the region 
containing the measured X-ray flux were determined directly from 
ROSAT  All Sky Survey (RASS).
All clusters have spectroscopically measured redshifts. 
The X-ray temperature was derived from the $L_X-T_X$ relation 
of White, Jones \& Forman (1997). The central electron densities and core 
radii were derived by fitting to the RASS data a spherically symmetric isothermal
$\beta$ model (Cavaliere \& Fusco-Femiano 1976) convolved with the RASS point-spread 
function. The $\beta$ was fixed at the canonical value of $2/3$
to reduce the dependence of the $\beta$ model parameters with
the choice of radius over which the model is fit. These data
allows to compute the Comptonization parameter at the center of the cluster. 

Atrio-Barandela et al (2008) compared the TSZ predicted from the X-ray data with
the signal present in WMAP 3yr data and found it to be in good agreement
within the X-ray emitting region, where the $\beta$ model is a good
description of the electron distribution. In the cluster outskirts, the
TSZ signal was systematically higher than the measured value. The latter 
was consistent with the Komatsu \& Seljak (2002) profile, where baryons 
are in hydrostatic equilibrium within a dark matter halo well described 
by a Navarro-Frenk-White profile (hereafter NFW, Navarro, Frenk \& White 1997), 
as expected in the concordance $\Lambda$CDM model. More recently, 
Nagai et al (2007) proposed a scaled 3-dimensional electron pressure 
profile $p(x)=P_e(r)/P_{500}$ based on a generalizad NFW profile
\begin{equation}
p(x) = \frac{P_0}{(c_{500}x)^\gamma[1+(c_{500}x)^\alpha]^{(\beta-\gamma)/\alpha}},
\label{eq:universal_profile}
\end{equation}
where $(\gamma,\alpha,\beta)$ are the central, intermediate and outer slopes,
$c_{500}$ characterizes the gas concentration and $x=r/R_{500}$ is the radius at
which the average density of the cluster is 500 times the critical density.
Arnaud et al (2010) derived an average cluster pressure profile from
observations of a sample of 33 local ($z<0.2$) clusters, scaled by mass
and redshift with
\begin{equation}
[P_0,c_{500},\gamma,\alpha,\beta]=[8.403 h_{70}^{-3/2},1.177,0.3081,1.0510,5.4905]
\label{eq:Arn_parameters}
\end{equation}
Later, Plagge et al (2010) showed these parameters to be consistent with 
the SZ measurements of 15 massive X-ray clusters observed with the South 
Pole Telescope (Plagge et al 2010). We determine the scale $R_{500}$ using 
(B\"ohringer et al 2007)
\begin{equation}
R_{500}=\frac{(0.753\pm0.063)h^{-1} \rm{ \ Mpc}}{h(z)}  
\left(\frac{L_X}{10^{44}h^{-2}\rm{\ erg\ s}^{-1}}\right)^{0.228\pm 0.015}
\label{eq:r500}
\end{equation}

\begin{figure}
\centering
\epsfxsize=.495\textwidth \epsfbox{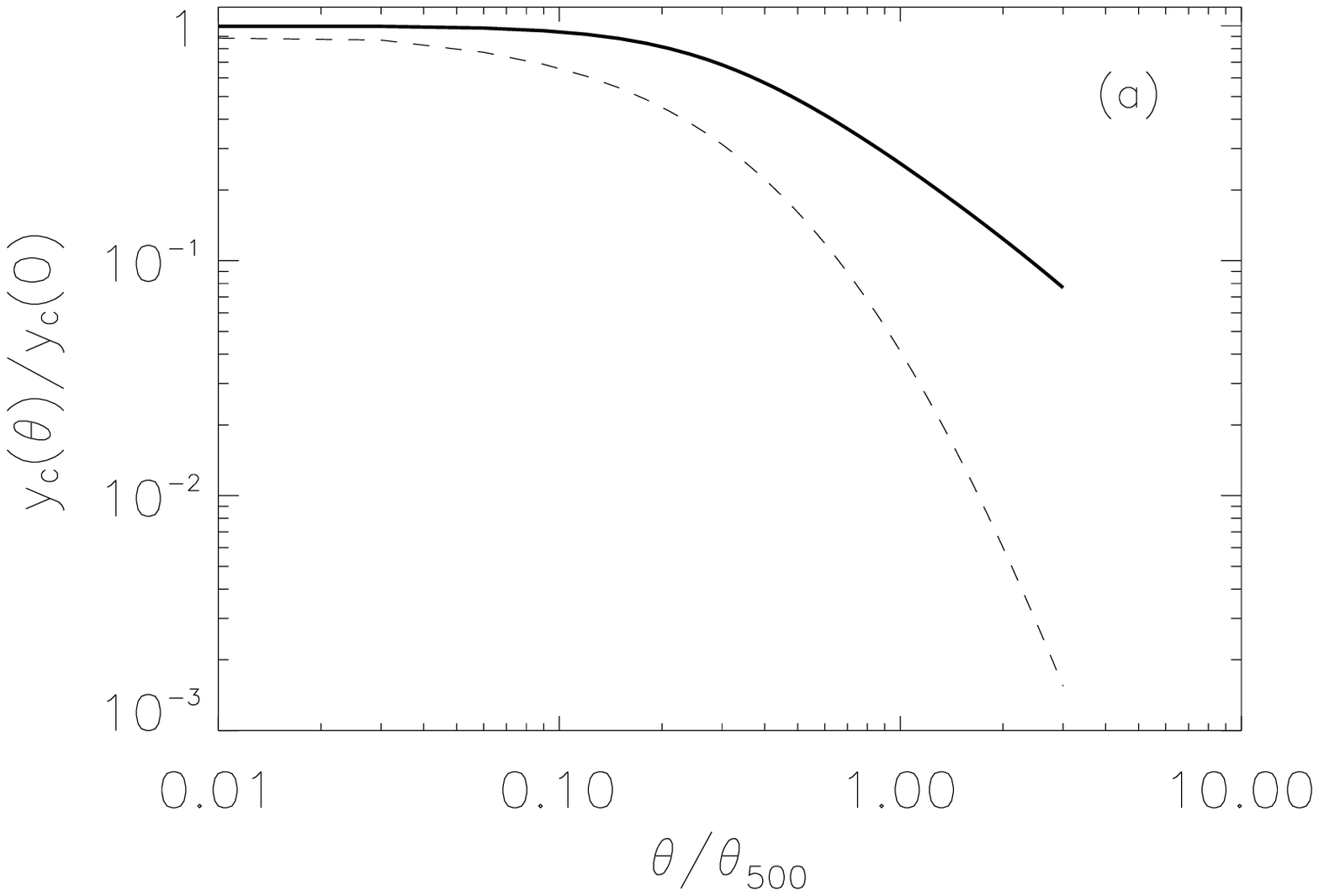}
\epsfxsize=.495\textwidth \epsfbox{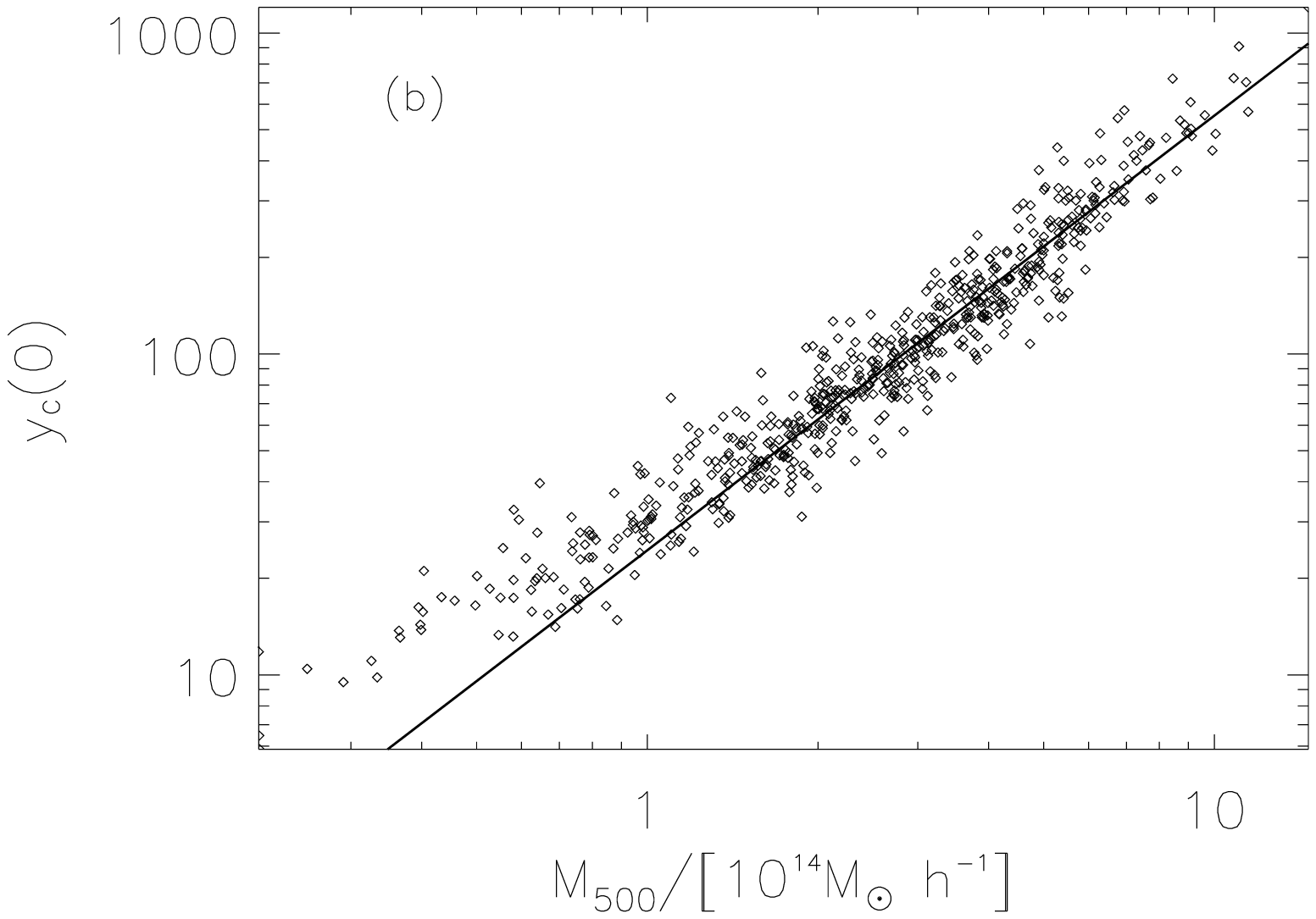}
\caption{(a) Pressure profile integrated along the line of sight of a cluster  with
$z=0.094$, of $M_{500}=2.4\times 10^{14}h^{-1}M_\odot$ and 
$R_{500}=750h^{-1}$Kpc convolved with the antenna of the 44GHz map.
Solid line corresponds to the $\beta=2/3$ model and the dashed line
to the universal pressure profile with the parameters of
eq.~\ref{eq:Arn_parameters}. Angles are expressed in units of $\theta_{500}$, 
the angle subtended by the radius $R_{500}$ at the position of the cluster.
(b) Central value of the Comptonization parameter for the clusters of
our proprietary sample, derived using the measured X-ray information.
The solid line correspond to the linear regression fit to the data.}
\label{fig3}
\end{figure}

To test the effect of the cluster profile on the final results, we construct 
y-maps from the X-ray cluster catalog (a) using the universal pressure profile 
of eq.~(\ref{eq:universal_profile}) with the parameters given in 
eq.~(\ref{eq:Arn_parameters}) and (b) using the isothermal $\beta=2/3$ model. 
The Comptonization parameter is
computed integrating the electron pressure profile along the line
of sight. Clusters are assumed to be spherically symmetric and extending up to 
$R_{200}$, the scale where the cluster overdensity reaches 200 times 
the critical density. To determine the effect of the cluster profile,
the central value of $y_c$ is assumed to be the same. Finally, the
cluster templates are convolved with the corresponding antenna beams (see Table 1).
In the Fig.~\ref{fig3}a we show the pressure profile integrated
along of line of sight for the $\beta=2/3$ (solid line)
and universal pressure (dashed line) profiles
convolved with the antenna of the 44GHz map. The cluster is located at
$z=0.094$, of $M_{500}=2.4\times 10^{14}h^{-1}M_\odot$ and
$R_{500}=746h^{-1}$Kpc. For illustration, in Fig.~\ref{fig3}b we plot 
the value of Comptonization parameter $y_c$ at the center of all the
clusters in our proprietary cluster catalog, derived
using the measured X-ray parameters,
as a function of cluster mass. The solid line
represents the linear regression fit to the data. The central Comptonization
parameter scales as: $y_c=24.5(M_{500}/10^{14}h^{-1}M_\odot)^{1.35}$.

\subsection{Y-map from simulated clusters. \label{sec:sim}}

As an alternative, we also use the low-redshift all-sky maps and the associated galaxy 
cluster catalogues of the {\it hydrodynamic} diffuse and kinetic SZ simulations included 
in the {\it pre-launch Planck Sky Model}. The simulations are fully described
in Dellabrouille et al (2011). 
The catalogues contain cluster positions, mass and radius for an overdensity 
contrast  of 200 times the critical density. The maps contain the integrated SZ 
signal up to $z\simeq 0.25$, computed from a combination of full hydrodynamic 
simulations using the box staking method described in Valente, da Silva 
\& Aghanim (2012). According to this method, the Universe around the observer 
is generated in concentric layers, each with a comoving thickness of 
$100h^{-1}$Mpc, using the outputs of hydrodynamic simulations with periodic 
boundary conditions. The light-cone integrations  
of the TSZ and KSZ signals are carried out using the formulae in
da Silva et al (2000) and (2001). A total of seven 
layers were constructed, up to z=0.25. The innermost layer includes the 
local constrained simulation of Dolag et al (2005), whereas all the 
other layers were produced from gas snapshots of the  $\Lambda$CDM simulation 
in De Boni et al (2010). Both these simulations include explicit treatment for 
gas cooling, heating by UV, star formation and feedback processes.

\begin{figure}
\epsfxsize=\textwidth\epsfbox{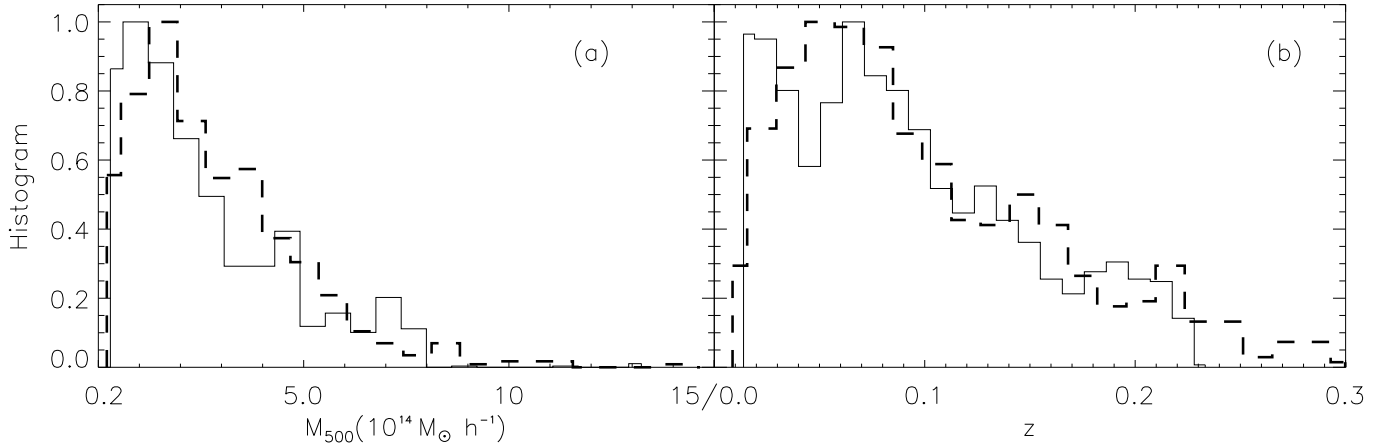} 
\caption{Mass (a) and redshift distribution (b) of
the catalog of X-ray selected clusters (dashed line) and of clusters selected
from the hydro-simulation described in Sec~\ref{sec:sim}.
For easier comparison, histograms are normalized to unity at the
maximum.}
\label{fig4}
\end{figure}

The y-map constructed from the X-ray selected clusters assumes clusters to be spherically
symmetric and relaxed while the TSZ and KSZ templates constructed from the hydrosimulation
contains clusters with different dynamical state (relaxed, merging systems, etc), 
shape and ellipticity. Also, since the latter are constructed integrating 
the signal along the line of sight, the projection effects due to low mass clusters 
and groups are included.  Therefore, these templates are
very well suited to study the effect of all these systematics and 
of the KSZ component in the determination of $\alpha$. For a more realistic 
comparison, we select 623 clusters from the simulation according to 
the measured selection function of the X-ray cluster sample. In Fig~\ref{fig4}a 
we plot the mass and in Fig~\ref{fig4}b the redshift distribution of
all clusters in our simulation (solid line) that fulfill the selection 
criteria. The dashed line shows the same distributions of the X-ray clusters.
For a better comparison, the histogram of largest amplitude
was normalized to unity. The main difference between the two samples is that 
there are 22 clusters in our proprietary cluster catalog that
have redshifts larger than $z=0.25$, the redshift of the last layer
constructed from the simulation. 

\subsection{Final Maps.} 

The y-maps described above are multiplied by
$G(\nu,\alpha=0)$ to generate TSZ templates and convolved with
the antenna beam. A KSZ template was added to
the hydrodynamical but not to the X-ray selected cluster template
since their peculiar velocity is not available. Noise maps were constructed 
assuming the Blue Book noise levels of Table~1.
When presenting our results, we shall demonstrate that the HFI frequencies
have the largest statistical power to constrain $\alpha$.
We model the noise as homogeneous and uncorrelated white noise since
at the frequencies $44-353$GHz the $1/f$ is both small and does not affect 
the angular scales subtended by clusters, $\ell\sim 500$ and above.

To take into account the two different component separation techniques, we 
carry out two different set of simulations: in (A) the CMB is removed in the TOI. 
Maps will only contain instrumental noise, TSZ and KSZ. In total, six
different maps, one for each frequency of Table~1 are simulated. In
(B) the 217GHz map is used to remove the intrinsic CMB. Then, only five difference
maps will be available for the analysis. Those maps 
will contain instrumental noise and TSZ, but the frequency
dependence of the TSZ effect changes. More realistic simulations would include 
foreground residuals, noise inhomogeneities with an $1/f$ component  that
can only be accurately model once the data becomes available. However, we do not
expect that our results obtained with our simplified maps will change with more
realistic simulations if noise inhomogeneities are uncorrelated with the 
cluster distribution. Then, we constructed two set of maps, according to 
the specific simulation.  In simulation (A) six maps, one for each 
frequency of Table~1, are constructed by adding noise to the TSZ and KSZ 
templates. The temperature anisotropy at 
each pixel is: $\Delta T_A(\nu)=y_cG(\nu,0)+\Delta T_{KSZ}\pm\sigma^A_{noise,\nu}$
In simulation (B) six maps are constructed adding cosmological CMB
signal and noise to the cluster templates. The 217GHz map is used
to subtract the cosmological and KSZ signals. Therefore, only
five different maps are available for the analysis.
We checked the final maps had a power spectrum that was a pure white noise, 
with a slightly larger variance $\sigma^B_{noise,\nu}$, sum of the
original map plus the noise of the degraded 217GHz map.
At each pixel, the temperature anisotropy is: 
$\Delta T_B(\nu)=y_cG_{[-217GHz]}(\nu,0)\pm\sigma^B_{noise,\nu}$.
We neglect relativistic corrections that are only significant for the
most massive clusters (Nozawa et al 1998).

\bigskip
\bigskip
\section{Data Processing.}

In both simulations A and B, we construct estimators using both the Ratio 
and the Fit methods. To simplify the notation, let the index
$I=(A,B)$ denote the type of simulation and let us redefine 
$G_A=G(\nu,\alpha)$, $R_A=R(\nu_1,\nu_2,\alpha)$,
$G_B=G_{[-217GHz]}(\nu,\alpha)$ and $R_B=R_{[-217GHz]}(\nu_1,\nu_2,\alpha)$.
At each cluster location, projection effects can yield contributions from
different redshifts altering the frequency dependence. To reduce the effect,
we will take averages over the cluster extent. The temperature anisotropy is then
\begin{equation}
\langle\Delta T_I(\nu_1)\rangle= \bar{y}_cG_I(\nu_1)
\pm\sigma^I_{Noise,\nu_1}/\sqrt{N_{pix}} .
\label{eq:deltatsz}
\end{equation}
where $N_{pix}$ is the number of pixels occupied by the cluster. There will
be an extra KSZ component for simulation A.

\subsection{Ratio method.}

To estimate $\alpha$ we compute the likelihood 
\begin{equation}
-2\log{\cal L}=\sum_{\nu_1,\nu_2}\sum_{i=1}^{N_{cl}}\left[
\frac{\langle\Delta T_I(\nu_1)\rangle/\langle\Delta T_I(\nu_2)\rangle-
R_I(\nu_1,\nu_2,\alpha)}{\sigma^I_{ratio,i}} \right]^2
\label{eq:chisq_ratio}
\end{equation}
for different values of $\alpha$.  A few examples of 
$R_I(\nu_1,\nu_2,\alpha)$ are  plotted in Fig~\ref{fig1}a and~\ref{fig2}a.
In eq.~(\ref{eq:chisq_ratio}), we compute $\sigma^I_{ratio,i}$ for each 
cluster as the rms deviation of 1,000 simulations of the ratio 
$\langle\Delta T_I(\nu_1)\rangle/\langle\Delta T_I(\nu_2)\rangle$
where the TSZ component is held fixed to the actual value at the cluster
location and the noise is drawn from a gaussian distribution with
zero mean and variance $(\sigma^I_{noise,\nu})^2/N_{pix,i}$.
As discussed in Luzzi et al (2009) the distribution of ratios is
dominated by the error on the denominator. Therefore, in our simulations
type (A) we exclude the 217GHz channel from the denominator, where 
the TSZ signal is null, to minimize the bias. 

\subsection{Frequency fit method.\label{sec:fitmethod}}

Alternatively, we can fit the TSZ signal of each cluster to the spectral
dependence of Fig~\ref{fig1}b and \ref{fig2}b. Similarly, the
likelihood function is 
\begin{equation}
-2\log{\cal L}=\sum_\nu\sum_{i=1}^{N_{cl}}
\left[\frac{\langle\Delta T_I(\nu)\rangle-\bar{y}_cG_I(\nu,\alpha)}
{\sigma^I_{noise,\nu,i}}\right]^2 ,
\label{eq:chisq_fit}
\end{equation}
where $\sigma^I_{noise,\nu,i}=\sigma^I_{noise,\nu}/\sqrt{N_{pix,i}}$.

\begin{figure}
\centering 
\epsfxsize=0.495\textwidth \epsfbox{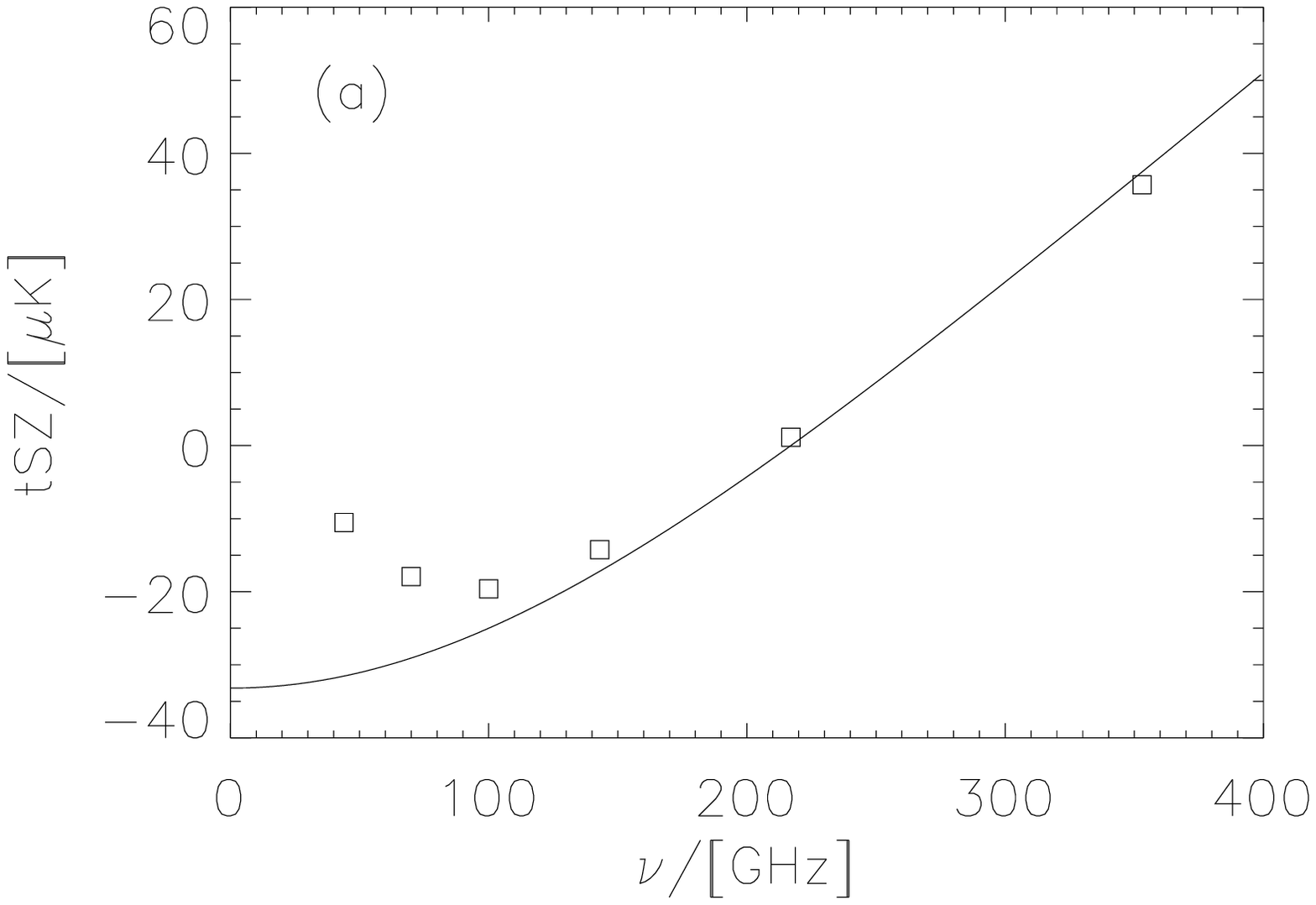}
\epsfxsize=0.495\textwidth \epsfbox{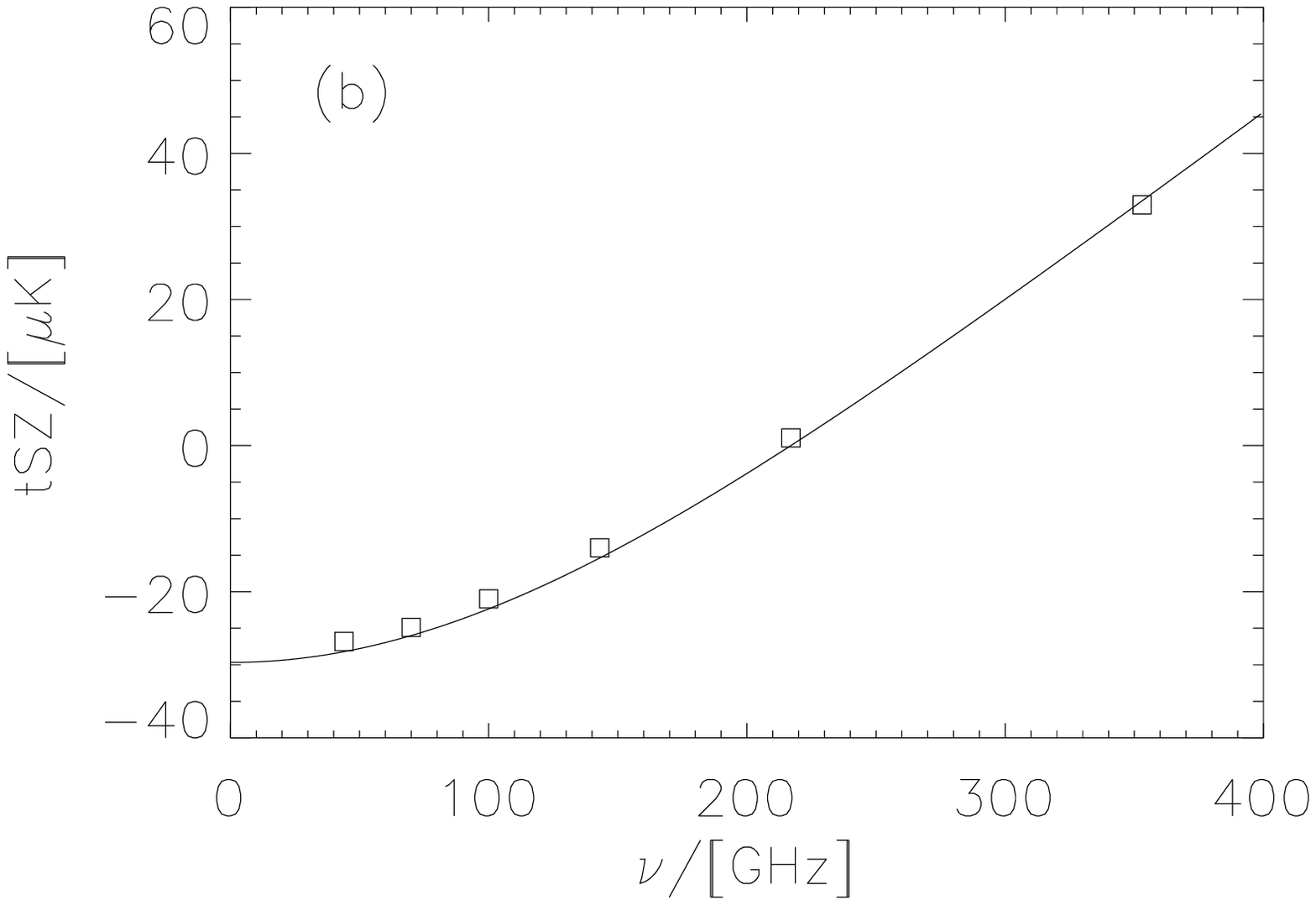}
\caption{Effect of the beam dilution on the spectral dependence of the
TSZ effect. Open squares represent the amplitude of the cluster
TSZ effect at different frequencies, averaged over a disk of extent 
$2\theta_{500}$. The solid line represents the TSZ scaling $G_(\nu,0)$.
Panel (a) corresponds to a cluster of mass $M_{500}=3.64\times 10^{14}h^{-1}M_\odot$
at redshift $z=0.218$, that subtends an angle $\theta_{500}=9.4'$.
Panel (b) corresponds to $M_{500}=7.7\times 10^{14}h^{-1}M_\odot$
located at redshift $z=0.058$, subtending an angle $\theta_{500}=42'$.}
\label{fig:convolved}
\end{figure}

This method requires an independent estimate of $y_c$, introducing another
complication. As is indicated in Table~1, different frequencies have different 
resolutions. The cluster anisotropies are diluted by the antenna beam and 
the TSZ signal does no scale as $G_I(\nu,0)$. As an example, in 
Fig~\ref{fig:convolved} open squares represent the average TSZ amplitude
on the 5 difference maps $\Delta T_{[\nu,-217GHz]}$; the solid line represent their 
frequency dependence $G_{[-217GHz]}(\nu,0)$ of Fig~\ref{fig2}b. Due to the antenna, 
the measured TSZ signal of the clusters differs from the expected scaling.  
The effect is most noticeable at 44GHz since this channel has the smallest 
resolution. The amplitude of the effect depends on the cluster profile 
and angular extent {\it but does not depend on the scaling of the TSZ signal
with redshift, $G(\nu,\alpha)$}. Fig~\ref{fig:convolved}a corresponds to a cluster 
at redshift $z=0.218$, with mass $M_{500}=3.64\times10^{14}M_\odot/h$ and size $9.4'$
while in  Fig~\ref{fig:convolved}b the cluster is located at $z=0.058$ 
with mass $M_{500}=7.7\times10^{14}M_\odot/h$ and size $42'$.

\begin{figure}
\centering 
\epsfxsize=0.495\textwidth \epsfbox{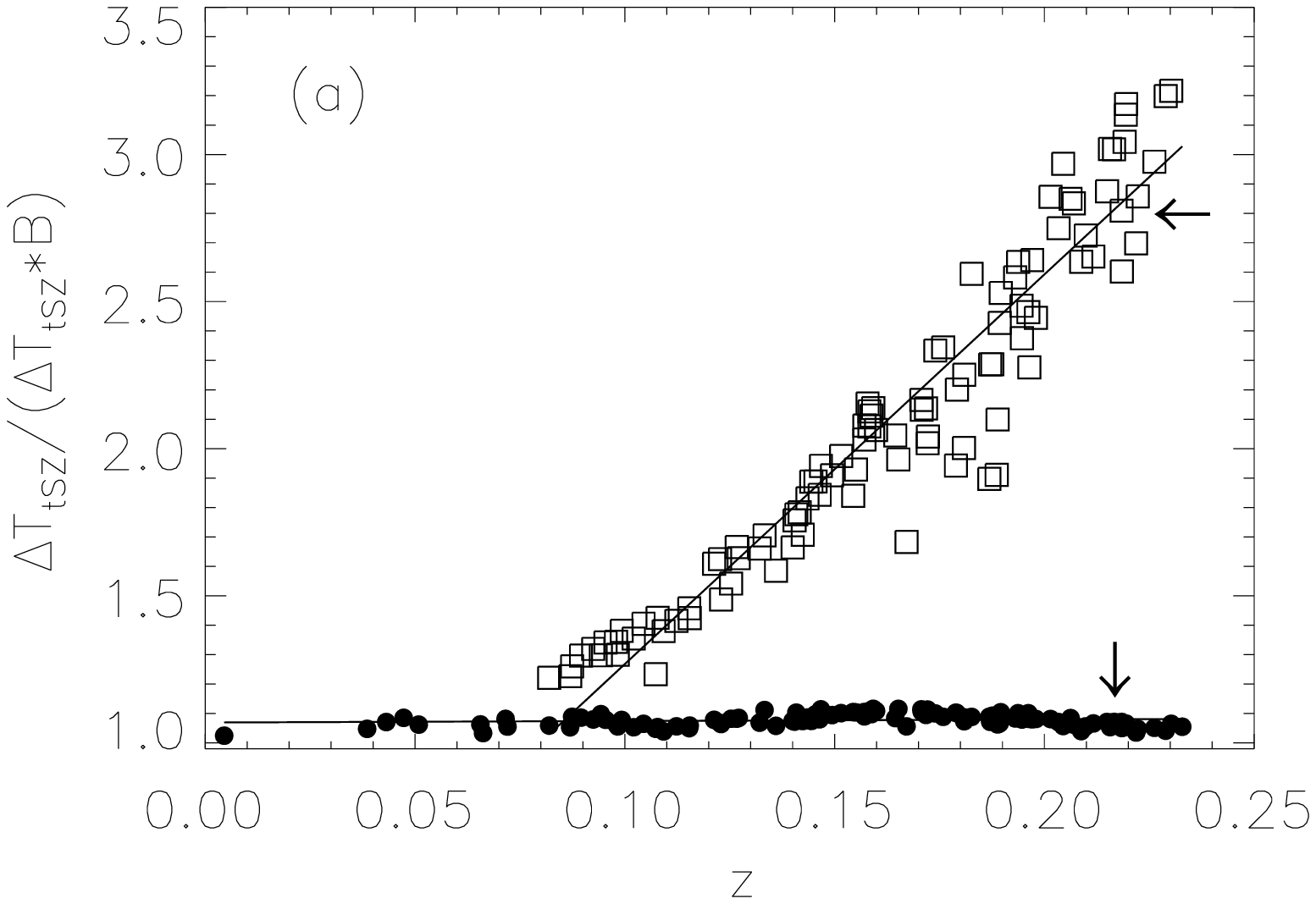}
\epsfxsize=0.495\textwidth \epsfbox{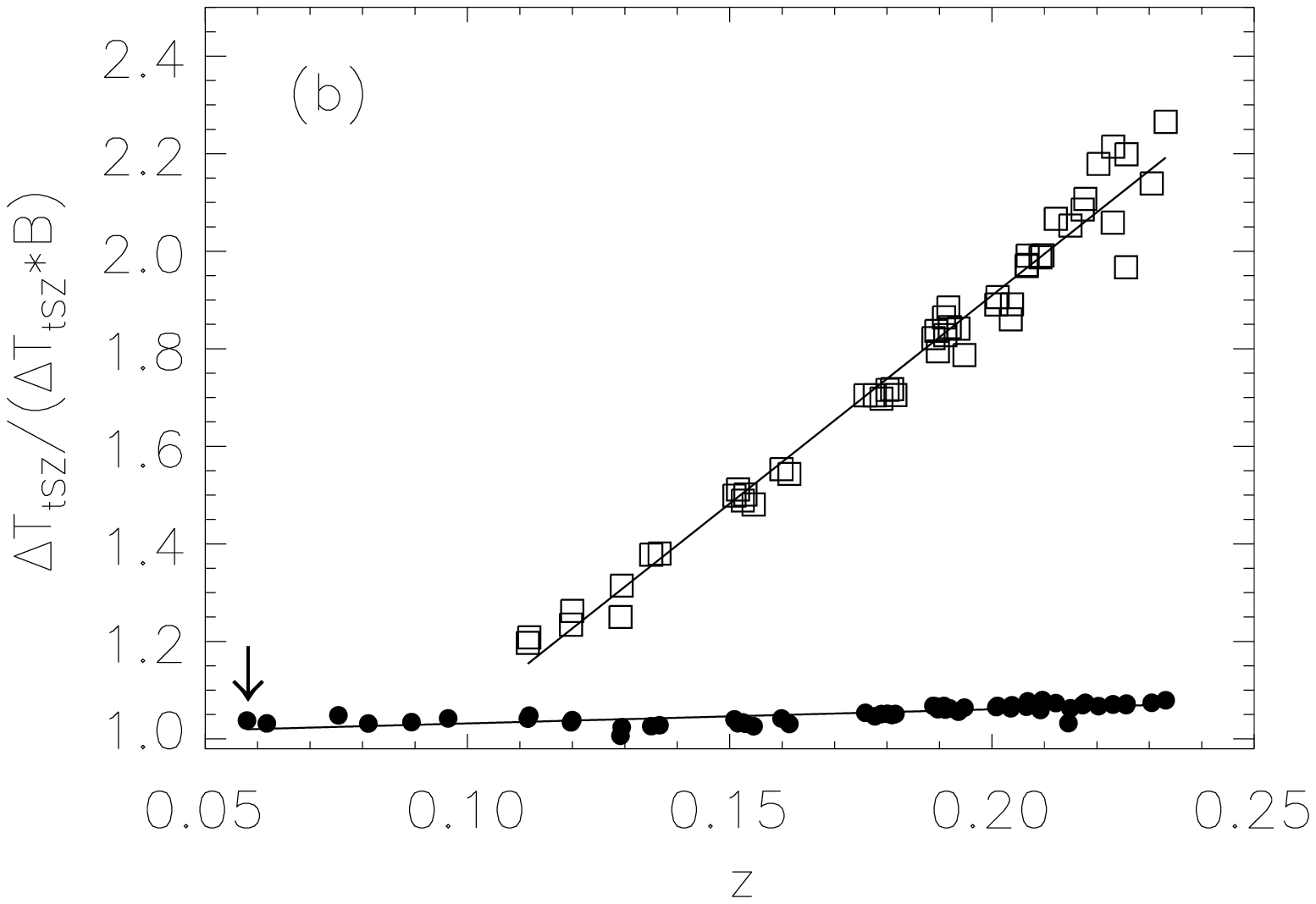}
\caption{ (a) Deconvolution factor for clusters in the mass range
$M_{500}=5-6\times 10^{14}h^{-1}M_\odot$ and (b) for clusters
with $M_{500}\ge 1\times 10^{15}h^{-1}M_\odot$. Solid black
circles represent the deconvolution factor for the 353GHz channel
and open squares for the 44GHz channel. All clusters are resolved
at 353 GHz but, for simplicity, at 44GHz only the fraction
of unresolved clusters is shown. Arrows indicate the deconvolution
factor of the clusters of Fig.~\ref{fig:convolved}.}
\label{fig:resolved}
\end{figure}

For clusters drawn from a simulation, its size, ellipticity and profile 
are known exactly. For such clusters, the deconvolution
factor $F$ can be determined exactly by comparing the average 
Comptonization parameter before ($\langle{y}_c\rangle$) and 
after $\langle y_c*B(\nu)\rangle$ convolving with the antenna beam $B(\nu)$:
$F=\langle{y}_c\rangle/\langle y_c*B(\nu)\rangle$. This factor would be different for 
resolved and unresolved clusters and would depend on the cluster profile and redshift.
For X-ray selected clusters, we know the Comptonization parameter in the cluster
cores but their pressure profile has not being measured.
This will introduce an extra uncertainty when comparing the measured TSZ
effect with the theoretical prediction. 
 For illustration, in Fig~\ref{fig:resolved}a we represent 
$F$ for a sample of 110 clusters in the mass range $M_{500}=5-6\times 
10^{14}M_\odot/h$. In Fig~\ref{fig:resolved}b, we plot the deconvolution factors
for all clusters in our simulation with masses $M_{500}\ge 10^{15}M_\odot/h$.
Solid black circles correspond to the 353GHz frequency and open squares to 44GHz. 
All clusters are resolved at 353GHz. At 44GHz clusters with redshift $z\ge 0.08$
are unresolved. For clarity the clusters at lower redshift, that would be resolved,
are not shown.

In Fig~\ref{fig:resolved} the solid straight lines correspond to
the linear regression fit to the deconvolution factor
for each cluster mass range and channel. Arrows indicate the deconvolution
factor of the clusters plotted in Fig~\ref{fig:convolved}a,b. 
If for each redshift, frequency and mass range, $F_{lin}$ is the deconvolution
factor estimated by the linear regression, $F$ the true deconvolution
factor and $\Delta F$ is the rms dispersion
of the true deconvolution values $F$ around $F_{lin}$ for each mass 
bin and antenna, then $F=F_{lin}\pm\Delta F$. If 
for a real cluster we use $F_{lin}$ instead of the (unknown) true factor
$F$, the deconvolved signal $(\Delta T_{TSZ}*B)F_{lin}$ would differ from the
the true signal $\Delta T_{TSZ}$ by an amount $(\Delta T_{TSZ}*B)\Delta F$.
This uncertainty is uncorrelated with the instrumental noise at the cluster 
location and can be included in the Likelihood analysis of
eq.~\ref{eq:chisq_fit} by adding it in quadrature with the instrumental noise:
$\sigma_{tot,i}^2=\sigma_{noise,i}^2+[(\Delta T_{TSZ}*B)\Delta F]^2$. 
On the other hand, the deconvolution coefficient does not scale linearly with
redshift, and $F_{lin}$ underestimates the true deconvolution factor $F$
especially at high redshifts, potentially biasing our estimation of $\alpha$.
We used the y-map computed with clusters drawn from a numerical simulation
to compute the deconvolution factors $F_{lin}$ and its uncertainty $\Delta F$
in three mass bins of equal number of clusters: $M_{500}\le 2\times
10^{14}M_\odot/h$, $M_{500}=2-3.6\times 10^{14}M_\odot/h$ and
$M_{500}\ge\times 3.6\times 10^{14}M_\odot/h$.
The deconvolution factors, that were different for resolved and unresolved
clusters, were used to deconvolve the templates of simulated and of X-ray clusters.

\bigskip
\bigskip
\section{Results and Discussion}

We first tested the ratio and fit methods using the template constructed from
simulations, as described in Sec.~\ref{sec:sim}. The template contained a 
subset of $623$ clusters distributed in mass and redshift according to the 
cluster catalog selection function (see Fig~\ref{fig4}) and included 
both TSZ and KSZ components. Second, we repeated the analysis with the template
of X-ray selected clusters. No KSZ contribution was added in this case.
The first and most important conclusion is that we found no significant
differences from the results computed using both templates, implying that
the effect of KSZ, cluster dynamical state and 
deviations from spherical symmetry are averaged out over such a large 
cluster sample, effects that were important when analyzing observations 
of just a few clusters (Battistelli et al 2002, Luzzi et al 2009).

With respect to the method of analysis, the ratio method performs differently
if the CMB is removed in the TOI (method A) or in the final map (method B)
but the differences are small for the fit method. For the ratio method,
we will only present results obtained with the simulated cluster template
and simulation type (A) (Fig~\ref{fig7}). For the fit method, the results presented 
are only using the template constructed from X-ray selected clusters
and simulation type (B) (Fig~\ref{fig8}) and we will discuss all the other
cases. To test the importance of the different contributions,
we define three mass bins $M_{500}=([<0.192],[0.192-0.365],[>0.36])
\times 10^{15} M_\odot/h$ of equal number ($\sim 208$) of clusters,
and three redshift bins $z=([<0.11],[0.11-0.17],[>0.17])$ with mean 
redshift $\langle z\rangle=(0.08,0.14,0.20)$, also with the same number
of clusters. We computed the likelihood 
(eqs.~[\ref{eq:chisq_ratio}] and [\ref{eq:chisq_fit}])
for the different mass and redshift bins and different frequencies to 
determine the data subset with the  largest statistical power.
To compute likelihoods, we subdivide the interval $\alpha=[-1,1]$ 
in 2001 steps. We perform 1,000 Monte-Carlo simulations for each
cluster template, simulation type and method.

\subsection{Ratio method.}

\begin{figure}
\centering 
\epsfxsize=0.95\textwidth \epsfbox{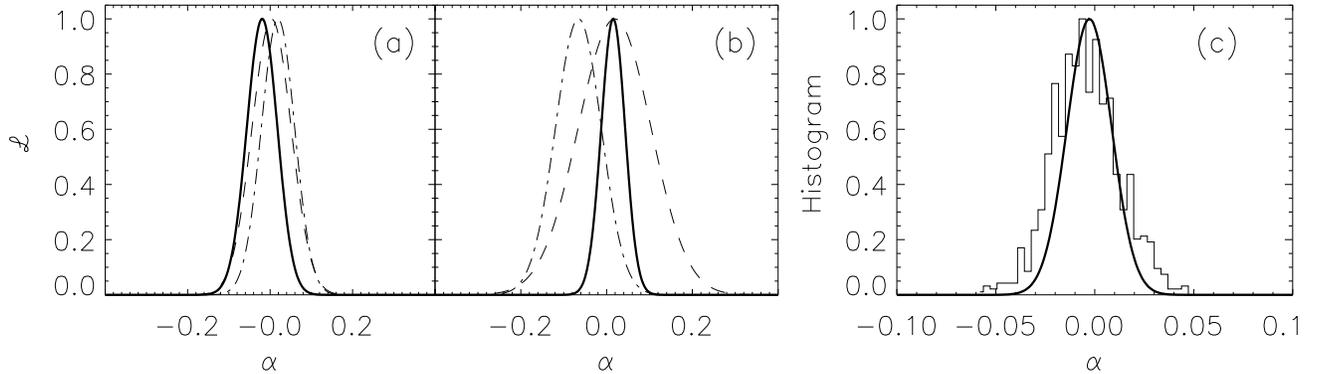}
\caption{Ratio method with simulation A. Likelihood function for subsamples of 
623 clusters distributed (a) in three redshift bins 
$z=([<0.11],[0.11-0.17],[>0.17])$ of equal number of clusters and
(b) in three mass bins $M_{500} = ([\le 1.9],[0.19−0.37],[\ge 3.7])
\times 10^{14} h^{-1}M_\odot$ also with the same number of clusters.
In both plots, dashed, dot-dashed and solid lines correspond to low,
intermediate and high redshift/mass bins. 
(c) Histograms of the value of $\alpha$ derived from 1,000 simulations,
arbitrarily normalized to unity. The smooth solid line
represents is a gaussian fit to the data.
}
\label{fig7}
\end{figure}

Our results indicate that the ratio method is strongly biased.
The likelihood function is dominated by the few clusters were the TSZ signal
is erased by the noise so $\langle \Delta T(\nu_2)\rangle\sim 0$ 
in the denominator of eq.~(\ref{eq:chisq_ratio}), biasing the results
towards negative values of $\alpha$. We found this bias is 
removed by rejecting all ratios where the denominator is smaller
than $0.2\sigma^A_{noise}$. This constrain rejects 5\% of the data
but reduces the bias to insignificant levels. 
In Fig.~\ref{fig7} we present the likelihood function of a single 
simulation randomly selected of our ensemble of 1,000 simulations. 
In Fig.~\ref{fig7}a we represent the likelihood for 
clusters within the three redshift bins given above, marginalized
over  cluster mass and 15 frequency ratios. Dashed, dot-dashed and solid lines
correspond to the lower, intermediate and high redshift bins.
In Fig.~\ref{fig7}b we present the results binning clusters
according to mass. Dashed, dot-dashed and solid correspond
to lower, intermediate and high mass bin. As expected,
the most massive clusters dominate the likelihood. The final
value obtained in this single simulation is $\alpha=0.0\pm 0.012$.
In Fig.~\ref{fig7}c we plot the histogram distribution of
the  $\alpha$ values measured in 1,000  simulations.
Solid line represents the results for simulation type A
where $\langle{\alpha}\rangle=-0.003\pm 0.011$. Our pipeline is marginally 
biased since the mean of our simulations $\langle{\alpha}\rangle=-0.003$
differs from $\alpha=0$ by more than $\sigma_{\alpha}/\sqrt{N_{sim}}=3.5\times 10^{-4}$.
For comparison, in Fig.~\ref{fig7}c the dashed line corresponds to 
simulation type B. The value of $\alpha$ averaged over all simulations 
is $\langle{\alpha}\rangle=-0.19\pm 0.18$. In this case, the larger uncertainty 
reflects the weak dependence of the ratio method with $\alpha$, as shown in
Fig~\ref{fig2}a. Finally, we also checked that ratios using lower resolution
maps have less statistical power to constrain $\alpha$.
In retrospect, this justifies neglecting the 30, 545 and 857GHz channels,
with large noise levels or (1/f) contributions, that would complicate
the analysis without adding more information. 

\subsection{Fit method.}

\begin{figure}
\centering 
\epsfxsize=0.95\textwidth \epsfbox{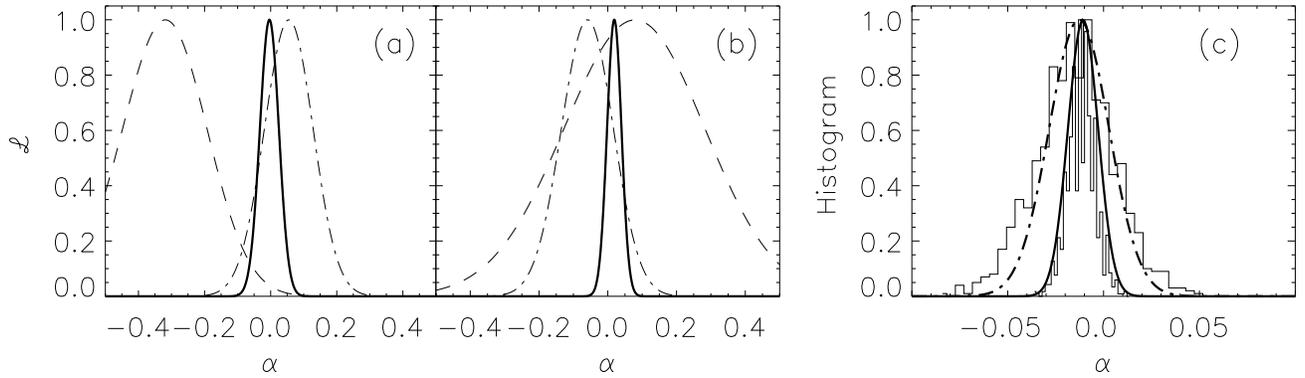}
\caption{Frequency Fit method with simulation B. The cluster template
was constructed using the universal pressure profile of 
eq.~\ref{eq:universal_profile}, with the parameters given in 
eq.~\ref{eq:Arn_parameters}. (a) Likelihoods for three
different frequencies: 44GHz (dashed), 100 GHz (solid)
and 343GHz (dot-dashed line) and (b) for three mass bins:
dashed, dot-dashed and solid lines correspond to the mass intervals 
$M_{500} = ([\le 1.9],[0.19−0.37],[\ge 3.7])\times 10^{14} h^{-1}M_\odot$.
(c) Histograms of the value of $\alpha$ derived from 1,000 simulations,
arbitrarily normalized to unity. The dot-dashed 
and solid lines correspond to the gaussian fits to the histograms of
y-maps constructed with the universal and $\beta=2/3$ profiles, respectively.
}
\label{fig8}
\end{figure}

The results of the fit method using a template of X-ray clusters and
simulations type B are presented in Fig.~\ref{fig8}.
The cluster template was constructed using the universal pressure profile of
eq.~\ref{eq:universal_profile}, with the parameters given in
eq.~\ref{eq:Arn_parameters}. In (a) we plot the likelihood function
for three different frequencies: 44GHz (dashed), 100 GHz (solid)
and 343GHz (dot-dashed line). The figure shows that the
100GHz channel is the most restrictive of the three. Like for the 
ratio method, the final
likelihood is dominated by the channels that have high resolution
and low noise, in this case the 100 and 143GHz channels. 
In Fig.~\ref{fig8}b we represent the likelihood 
for the three mass bins given above and marginalized over frequencies;
dashed, dot-dashed and solid lines correspond to the low, intermediate
and high mass intervals. The signal is dominated by the most massive
clusters that, on a flux limited sample, are on average at high 
redshift than the lower and intermediate mass samples.
For this particular realization, the estimated value is $\alpha=0.013\pm 0.020$,
compatible with adiabatic evolution at the $1\sigma$ level.

In Fig.~\ref{fig8}c we represent the histograms of 1,000 
simulations together with their linear fits for 
cluster templates constructed using the universal pressure profile
(dot-dashed line) and the $\beta=2/3$ (solid line). 
The mean and rms dispersion of the estimated values are
$\langle{\alpha}\rangle=-0.013\pm 0.016$ for the universal profile and
$\langle{\alpha}\rangle=0.003\pm 0.008$ for the $\beta$-model profile. 
When all cluster properties are identical, the TSZ integrated over the
cluster extent will be larger for the $\beta$-model than for the universal
profile (see Fig~\ref{fig3}a) so it must constrain $\alpha$ better, 
as shown.  We also carried out 1,000 simulations using method A, that does 
not include the intrinsic CMB signal, with the hydrodynamical template, that
contains the KSZ component. The result was $\langle{\alpha}\rangle=-0.008\pm 0.015$,
identical to the result with the method B above. Therefore, in the fit method
is not so important to have maps with the cosmological signal removed
as, for example, in the ratio method. Using the 217GHz map to remove the
intrinsic CMB signal alters the TSZ frequency dependence, but the TSZ signal
is still strongly dependent with redshift (see Figs.~\ref{fig1}b and \ref{fig2}b)
what is not the case in the ratio method (compare Figs.~\ref{fig1}a and \ref{fig2}a).

Let us remark that the rms dispersion of $\alpha$ on 1,000 simulations,
$\sigma_\alpha$, is very similar to the error on $\alpha$ 
in one single realization, both in the ratio and in the fit method, 
indicating that our pipelines are efficient. The results obtained 
using y-maps constructed with clusters drawn from a hydrodynamical
simulation or from a catalog of X-ray selected clusters show no significant
differences. In the hydro-simulation, the y-map integrates the SZ signal 
up to $z\simeq 0.25$ and contains all the projection effects up to that redshift, 
not included in the
X-ray selected clusters template, we can conclude that projection effects play no
significant role. This can be understood in the light of the results presented
in Fig~\ref{fig8}b; the full likelihood is dominated by the most massive clusters
for which projection effects are not significant (Valente et al 2012).

\bigskip
\bigskip
\section{Conclusions.}

Planck offers an excellent opportunity to constrain
the evolution history of the CMB blackbody temperature with better precision
than quasar excitation lines using currently available X-ray cluster
catalogs. We have found 
that taking the ratio of temperature anisotropies at different frequencies
is strongly biased but this bias can be corrected by rejecting all ratios where 
the denominator is much smaller than the noise. Fitting the frequency 
dependence provides an equally reliable estimator with
different systematics but requires both an independent determination 
of the Comptonization parameter and deconvolution of the antenna beam. 
The latter can not be done exactly if the cluster pressure profile is not 
known precisely. We have shown that deconvolution using linear fits
introduces an error that can be easily incorporated into the analysis. 

We have considered two possible methods to remove foregrounds and the 
cosmological CMB signal and the KSZ contribution:
the cosmological signal is removed in the TOI and 
the 217GHz map is used to remove the cosmological and KSZ signal exactly.
We have carried out simulations of both methods to investigate 
the differences on the final results. We have shown that the ratio method 
performs rather well if the cosmological CMB signal is clean in the
TOI but very badly otherwise. The fit method performs equally well 
in both data sets, giving results that are only marginally biased. 
With both methods, massive clusters and the high resolution/low noise 
channels have the largest statistical power to constrain $\alpha$.
We have used a proprietary cluster catalog that contains
spectroscopic redshifts and all the required X-ray
information to estimate the accuracy that would be achieved with Planck data. 
We forecast that the final uncertainty will be about $0.011-0.016$ a factor 
2-3 better than those obtained from quasar spectra by Noterdaeme et al (2010),
depending on what type of Planck data becomes publicly available.

Since our catalog is restricted to 
clusters with $z\le 0.3$, we have not extended our analysis beyond that
redshift. Planck has already detected around 200 clusters with a $S/N\ge 10$,
one at $z\simeq 0.94$ with $M_{500}\simeq 8\times 10^{14} M_\odot$. Adding
more clusters with current or future experiments will help to detect
possible deviations from adiabatic evolution, specially if 
clusters are of higher mass and are at a higher redshift
like the one recently reported in Planck Collaboration 2011c. 
Once all PLANCK and South Pole cluster candidates have been observed
on the X-ray and their redshift determined, the measurements proposed 
will provide much stronger constraints on non-adiabatic evolution,
than those quoted here.

\bigskip
\bigskip
\section{Acknowledgements}

This work was done in the context of the FCT/MICINN cooperation grant
'Cosmology and Fundamental Physics with the Sunyaev-Zel'dovich Effect'
AIC10-D-000443, with
additional support from project PTDC/FIS/111725/2009 from FCT, Portugal
and FIS2009-07238 and CSD 2007-00050 from the Ministerio de Educaci\'on y 
Ciencia, Spain. The work of CM is funded by a Ci\^encia2007 Research Contract,
funded by FCT/MCTES (Portugal) and POPH/FSE (EC).

\end{document}